\documentclass[12pt]{article}
\usepackage{amsmath,amssymb,amsbsy,amsthm,latexsym}

\advance\topmargin by -\headheight
\advance\topmargin by -\headsep
\evensidemargin=\oddsidemargin
\newcommand{\A}{\mathcal{A}}        
\renewcommand{\a}{\alpha}           
\renewcommand{\AA}{\mathbb{A}}      
\newcommand{\Ahat}{\Hat{A}}
\newcommand{\as}{\quad\text{as}\enspace} 
\renewcommand{\b}{\beta}            
\newcommand{\C}{\mathbb{C}}         
\newcommand{\D}{\mathcal{D}}        
\newcommand{\del}{\partial}         
\newcommand{\delslash}{\partial\mkern-9mu/} 
\newcommand{\Dl}{\Delta}            
\newcommand{\eps}{\varepsilon}      
\newcommand{\Ga}{\Gamma}            
\newcommand{\ga}{\gamma}            
\DeclareMathOperator{\GeV}{GeV}     
\renewcommand{\H}{\mathcal{H}}      
\renewcommand{\Hat}[1]{\widehat{#1}}  
\newcommand{\HH}{\mathbb{H}}        
\DeclareMathOperator{\Hom}{Hom}     
\DeclareMathOperator{\id}{id}       
\newcommand{\K}{\mathcal{K}}        

\newcommand{\La}{\Lambda}           
\newcommand{\la}{\lambda}           
\newcommand{\lab}{\boldsymbol{\lambda}} 
\newcommand{\lahat}{\hat\la}        
\newcommand{\Onda}[1]{\widetilde{#1}} 
\newcommand{\ox}{\otimes}           
\newcommand{\pd}[2]{\frac{\partial#1}{\partial#2}} 
\newcommand{\R}{\mathbb{R}}         
\newcommand{\Rbar}{\overline{R}}    
\newcommand{\sepword}[1]{\quad\text{#1}\quad} 
\newcommand{\set}[1]{\{\,#1\,\}}     
\renewcommand{\SS}{\mathcal{S}}     
\DeclareMathOperator{\Str}{Str}     
\newcommand{\stroke}{\mathbin\vert} 
\newcommand{\sul}{\mathfrak{su}}    
\newcommand{\taub}{\boldsymbol{\tau}} 
\DeclareMathOperator{\TeV}{TeV}     
\newcommand{\tihalf}{\tfrac{i}{2}}  
\def\top{\mathrm{top}}              
\newcommand{\tquarter}{\tfrac{1}{4}}  
\DeclareMathOperator{\Tr}{Tr}       
\DeclareMathOperator{\tr}{tr}       
\newcommand{\ul}{\mathfrak{u}}      
\newcommand{\x}{\times}             
\newcommand{\1}{\mathbf{1}}         
\newcommand{\7}{\dagger}            
\renewcommand{\:}{\colon}           
\def\<#1,#2>{\langle#1\stroke#2\rangle} 
\def\(#1,#2){(#1\stroke#2)}

\newcommand{\sunset}{%
\parbox{15mm}{\begin{picture}(20,10)
\put(0,5){\line(1,0){10}}
\put(20,5){\circle{20}}
\put(10,5){\line(1,0){30}}
\end{picture}}
}
\newcommand{\iice}{%
\parbox{21mm}{\begin{picture}(20,10)
\put(0,5){\line(1,0){10}}
\put(20,5){\circle{20}}
\put(40.2,5){\circle{20}}
\put(30.2,5){\oval(40.5,40)[bl]}
\put(30.2,5){\oval(40,40)[br]}
\put(50.2,5){\line(1,0){10}}
\end{picture}}
}
\newcommand{\sweet}{%
\parbox{10mm}{\begin{picture}(20,10)
\put(0,5){\line(-1,2){5}}
\put(0,5){\line(-1,-2){5}}
\put(10,5){\circle{20}}
\put(20,5){\line(1,2){5}}
\put(20,5){\line(1,-2){5}}
\end{picture}}
}
\newcommand{\dsweet}{%
\parbox{15mm}{\begin{picture}(20,10)
\put(0,5){\line(-1,2){5}}
\put(0,5){\line(-1,-2){5}}
\put(10,5){\circle{20}}
\put(30.2,5){\circle{20}}
\put(40.2,5){\line(1,2){5}}
\put(40.2,5){\line(1,-2){5}}
\end{picture}}
}
\newcommand{\tsweet}{%
\parbox{23mm}{\begin{picture}(20,10)
\put(0,5){\line(-1,2){5}}
\put(0,5){\line(-1,-2){5}}
\put(10,5){\circle{20}}
\put(30.2,5){\circle{20}}
\put(50.4,5){\circle{20}}
\put(60.2,5){\line(1,2){5}}
\put(60.2,5){\line(1,-2){5}}
\end{picture}}
}
\newcommand{\kg}{%
\parbox{19mm}{\begin{picture}(20,10)
\put(0,5){\line(-1,2){5}}
\put(0,5){\line(-1,-2){5}}
\put(10,5){\circle{20}}
\put(20,5){\line(2,1){30}}
\put(20,5){\line(2,-1){30}}
\qbezier(35,5)(35,10)(40,15)
\qbezier(35,5)(35,0)(40,-5)
\qbezier(45,5)(45,10)(40,15)
\qbezier(45,5)(45,0)(40,-5)
\end{picture}}
}
\newcommand{\pinterIV}{%
\parbox{16mm}{\begin{picture}(20,10)
\put(10,5){\line(-2,-1){10}}
\put(10,5){\line(-2,1){10}}
\put(10,5){\line(2,1){30}}
\put(10,5){\line(2,-1){30}}
\qbezier(25,5)(25,10)(30,15)
\qbezier(25,5)(25,0)(30,-5)
\qbezier(35,5)(35,10)(30,15)
\qbezier(35,5)(35,0)(30,-5)
\end{picture}}
}

\title{Noncommutative Geometry and Fundamental Interactions:
The First Ten Years}

\author{
Jos\'e M. Gracia-Bond\'{\i}a\\[1pc]
Departamento de F\'{\i}sica Te\'orica I, Universidad Complutense,\\
28040 Madrid, Spain\thanks{
Permanent address: 
Departamento de F\'{\i}sica, Universidad de Costa Rica,
2060 San Pedro de Montes de Oca, Costa Rica.}}

\begin{document}

\maketitle

\begin{abstract}
This is the full text of a survey talk for nonspecialists, delivered
at the 66th Annual Meeting of the German Physical Society in Leipzig,
March 2002. We have not taken pains to suppress the colloquial style.
References are given only insofar as they help to underline the points
made; this is not a full-blooded survey. The connection between
noncommutative field theory and string theory is mentioned, but
deemphasized. Contributions to noncommutative geometry made in Germany
are emphasized.
\end{abstract}

\section{Mathematical prehistory}

As a branch of mathematics, Noncommutative Geometry (NCG) is 20 years
old. It was precisely in Germany, at the Oberwolfach meeting in
September--October 1981, that Connes unveiled a ``homology of currents
for operator algebras''~\cite{ConnesObW} amounting to
\textit{differential analysis on noncommutative algebras}.

Geometric properties of noncommutative algebras had already been
worked out in detail for the noncommutative torus, then known as the
``irrational rotation $C^*$-algebra'', by Connes
himself~\cite{ConnesTorus}; whereupon the canonical trace on the torus
algebra gives a ``noncommutative integral''.

The related calculus, reformulated in the new language of cyclic
cohomology, was developed in detail in the foundational paper on
``Noncommutative Differential Geometry''~\cite{ConnesNCDiffG}, which
started to circulate in preprint form around Christmas, 1982. A high
point of this article is the so-called
Hochschild--Kostant--Rosenberg--Connes theorem~\cite{Polaris}. By
1983, Connes' approach to cyclic cohomology was essentially
complete~\cite{ConnesLambda}; his parallel work on foliations and
noncommutative integration led him to emphasize the role of Fredholm
modules in the NCG formalism. The Baum--Connes and Connes--Kasparov
conjectures were formulated in their primitive form in the early
eighties. Also pioneer work by Rieffel~\cite{RieffelRot} contributed
to shape noncommutative geometry in its early stages.

\smallskip

Noncommutative geometry, in essence, is an operator algebraic,
variational reformulation of the foundations of geometry, extending to
noncommutative spaces. NCG allows consideration of ``singular
spaces'', erasing the distinction between the continuous and the
discrete. Its main specific tools are Dirac operators, Fredholm
modules, the noncommutative integral, Hochschild and cyclic cohomology
of algebras, $C^*$-modules and Hopf algebras. NCG has many affinities 
with quantum field theory, and it is not unusual for noncommutative 
homological constructs to crop up in that context~\cite{Halley}.

On the mathematical side, NCG has had a vigorous development. Current
topics of interest include index theory and groupoids, mathematical
quantization, the Novikov and Baum--Connes conjectures and the
relation of the latter to the Langlands program, locally compact
quantum groups, the dressing of fermion
propagators in the framework of spectral triples, and the famous
Riemann hypothesis. In the mainstream of mathematics, the
noncommutative program needs no tribute. It is here to stay.

\section{The Interface with Physics}

We say interface because indeed, there is no question of
``application'' of NCG to physics, but rather of mutual intercourse.
In fact, the original use of noncommutative geometry in physics was a
model of epistemological humility. Instead of trying to
{\it derive\/} the laws of physics from some NCG construct, people
were trying to {\it learn}, from the mainstream physical theories,
what the noncommutative geometry of the world could be. By
``mainstream physical theories'' we essentially understand the
Standard Model (SM) of fundamental interactions.

The Standard Model, which has been been termed ``ugly'' by so many
physicists, including some of its fathers, is dramatically elegant and
beautiful in modern mathematics, in that its crucial concepts are
among the deepest and most powerful in noncommutative geometry. I am
talking about the concepts of gauge field and of chiral fermion.

Gauge fields are identical with connections, perhaps the most
important objects in the modern formulation of geometry. The algebraic
definition of linear connection is imported verbatim into NCG. Thinking
seriously about the space of all connections has been a very fruitful
idea that mathematicians have picked up from physicists~\cite{Woit}.

Chiral fermions are acted on by Dirac and Dirac--Weyl operators.
Perhaps it is not so well known that Dirac operators are a source of
NCG, where they give rise to ``fundamental classes''. (Any complex
spinor bundle on a smooth manifold gives rise to a generalized Dirac
operator $D$, whose sign operator $F = D|D|^{-1}$ determines a
Fredholm module; the natural equivalence class of these depends only
on the underlying spin$^c$-structure~\cite{HigsonR}. Since the
spin$^c$ structure determines, among other things, the orientation of
the manifold, this class is a finer invariant than the usual
fundamental class in homology.) It is precisely the role of the Dirac
operator both in the Standard Model and in NCG that made the
rapprochement between this branch of mathematics and fundamental
physics natural and unavoidable.

There have been two main lines for making sense of the SM in terms of
NCG: the one developed by Connes himself (and followers), and the one
related to the Lie superalgebra $\sul(2|1)$, mainly exploited by the
Mainz--Marseille group (Scheck, H\"au{\ss}ling, Paschke, Papadopoulos,
Coquereaux, Esposito-Far\`ese, Vaillant), more or less at the
same time.

Let us collect the gauge potentials appearing in the SM into a single
package of differential forms:
$$
\AA' = i(B,W,A),
$$
where
$$
B = -\tihalf g_1 \mathbf{B}_\mu \,dx^\mu, \quad
W = -\tihalf g_2\,\taub \cdot \mathbf{W}_\mu \,dx^\mu
\quad{\rm and}
\quad A = -\tihalf g_3\,\lab \cdot \mathbf{A}_\mu\,dx^\mu,
$$
with $\mathbf{B}$, $\mathbf{W}$ and $\mathbf{A}$ denoting respectively
the hypercharge, weak isospin and colour gauge potentials. From the
mathematical standpoint, $W$ is to be regarded as a 1-form with values
in the real field $\HH$ of quaternions. In other words, $\AA'$ is an
element of $\La^1(M) \ox \A_F$, where the noncommutative algebra
$\A_F := \C \oplus \HH \oplus M_3(\C)$, that we baptized the 
\textit{Eigenschaften\/} algebra~\cite{Cordelia}, plays the crucial
r\^ole.

Let us also collect all chiral fermion fields into a multiplet $\Psi$
and denote by $J$ the charge conjugation operation. Then the fermion 
kinetic term is rewritten as follows:
\begin{equation}
I(\Psi,\AA',J) = \<\Psi, (i \delslash + \AA' + J\AA' J^\7) \Psi>.
\label{eq:action1}
\end{equation}
Next we look at the Yukawa part of the SM Lagrangian. Use the
normalized Higgs doublet
$\Phi := \begin{pmatrix} \Phi_1\\ \Phi_2\end{pmatrix}
:= \sqrt2\,\phi/v$, where $\phi$ is a Higgs doublet with vacuum
expectation value~$v/\sqrt2$. We also need
$$
\Onda\Phi := \begin{pmatrix} -\bar\Phi_2 \\ \bar\Phi_1 \end{pmatrix}
= \begin{pmatrix} 0 &-1\\ 1 & 0 \end{pmatrix}
\begin{pmatrix} \bar\Phi_1 \\ \bar\Phi_2 \end{pmatrix}.
$$
It has been remarked that the Higgs should also be properly regarded
as a quaternion-valued field; this makes apparent the custodial
symmetry of the scalar sector of the SM, which is related to the
$\rho = 1$ tree level prediction for the low-energy neutral to charged
current interactions ratio.

Introduce then $q_\Phi = \begin{pmatrix} \bar\Phi_1 & \bar\Phi_2 \\
- \Phi_2 & \Phi_1 \end{pmatrix}$ and
$r := q_\Phi - \langle q_\Phi \rangle = q_\Phi - 1$ and write,
schematically for a right-left splitting of the fermion multiplets:
$$
\AA'' = \begin{pmatrix} & M^\7 r \\ r^\7 M & \end{pmatrix},
$$
where $M$ denotes the quark or lepton mass matrix (including the
Kobayashi--Maskawa or Maki--Nakagawa--Sakata parameters), as the case
might be. Just as the standard Dirac operator relates the left- and
right-handed spinor representations, the Yukawa operator $\D_F$ in the
space of internal degrees of freedom relates also the left- and
right-handed chiral sectors. The Yukawa terms for both particles and
antiparticles (for the first generation) are written
\begin{align*}
I(\Psi,\AA'',J) &:= \<\Psi, (\D_F + \AA'' + J\AA''J^\7) \Psi>
\\
&= \bar q_L \Phi\,m_d\,d_R + \bar q_L \Onda\Phi\,m_u\,u_R
+ q_R \bar \Phi\,\bar m_d\, \bar d_L
+ q_R \Onda{\bar\Phi}\,\bar m_u\, \bar u_L
\\
&\qquad
+ \bar\ell_L \Phi\,m_e\,e_R
+ \bar\ell_L \Onda\Phi\,m_\nu\,\nu_R
+ \ell_R \bar\Phi\, \bar m_e\, \bar e_L 
+ \ell_R  \Onda{\bar\Phi}\,\bar m_\nu\, \bar \nu_L
+ {\rm h.c.}
\end{align*}
Note that neutrinos are assumed to have only Dirac mass terms, and
that the hypercharges of $\Phi$, $\Onda\Phi$ are respectively $+1$
and~$-1$.

The combination of both constructions yields a Dirac--Yukawa operator
$\D = i\delslash \oplus \D_F$. With $\AA := \AA'\,\oplus\,\AA''$, the
\textit{whole} fermionic sector of the SM being recast in a form
similar to~\eqref{eq:action1}:
\begin{equation}
I(\Psi,\AA,J) = \<\Psi, (\D + \AA + J\AA J^\7) \Psi>.
\label{eq:action2}
\end{equation}

In keeping with the old Kaluza--Klein idea, we have combined the
ordinary gauge fields and the Higgs as entries of a
\textit{generalized gauge field}. The Yukawa terms come from the
minimal coupling recipe applied to the gauge field in the internal
space. Hence the ``two-leaves'' picture for the
Glashow--Weinberg--Salam model~\cite{Sirius, KastlerS} that contributed to make
intuitive the noncommutative vision of the SM. The Dirac--Yukawa
operator is seen to contain in NCG all the relevant information
pertaining to the~SM.

The construction was made in the realm of Euclidean field theory. I
believe Mario Paschke will explain to us in this Meeting the
(difficulties and) hopes for introducing time in the picture.

This Connes--Lott reconstruction of the SM gave rise to two
``predictions''. It is important to recall that, at the time they were
made, the top quark had not yet been seen, and the best estimates for
its mass clustered around $130 \GeV$! The NCG model sort of explains
why the masses of the top quark, the $W$ and $Z$ particles and the
Higgs particle should be of the same order, and gave right away
$$
m_\top \geq \sqrt{3} m_W \approx 139 \GeV.
$$
With a bit of renormalization group running~\cite{Orpheus}, it fell
right on the mark.

On the other hand, the ``prediction'' for the Higgs mass from Connes'
NCG has remained stubbornly around $200 \GeV$, which is too large for
current phenomenological prejudice.

\smallskip

The Mainz $\sul(2|1)$ model was based on the following idea: put
together the $(W,B)$ forms belonging to the $\sul(2) \x \ul(1)$ gauge
group into a $3 \x 3$ antihermitian matrix and fill up the lateral
columns with the Higgs field components:
$$
M = \begin{pmatrix}
W_{2\x2} & C_{2\x1} \\ D_{1\x2} & B_{1\x1}
\end{pmatrix}_{3\x3},
$$
with
$$
\Str M = \tr W - B = 0.
$$
This gives rise to the Lie superalgebra, and then we can look for its
representations and relate them to the fermion fields. The idea goes
back to~\cite{Neeman}, but its more modern proposers realized that the
superalgebra should not be gauged.

It turns out that the Mainz model gives as good a description as the
Connes--Lott machine of spontaneous symmetry breaking, and that the
known families of quarks and leptons fit very well with the (reducible
but indecomposable) $\sul(2|1)$ representations~\cite{Scheck}.

This model received less attention than Connes', perhaps because the
Dirac operator does not play the main role. However, the big prize
nowadays, the true pot of gold, lies in ``\textit{horizontal
symmetry}''; all the more so since we have come to realize that the
neutrino sector has a much richer structure which was found a few
years ago. The attraction of the Mainz model, or variants thereof,
over Connes', is that it gives some clues to the structure of the
fermion mass matrix; and so, that model is alive and well.

On this score, I offer a decidedly low-tech suggestion. Let us start
from the same idea of completing the $\sul(2) \x \ul(1)$ into a
$3 \x 3$ matrix. Go back to the ordinary Lie algebra context in mixing
the four Higgs field components with a $\ul(3)$ symmetry (it has
always been clear that $\sul(3)$ won't work). The extra scalar field
comes in handy for astrophysical purposes.

In other words, we recognize the basic fact from covariant wave
equations theory, and from the very structure of the SM that spin~0
and spin~1 fields always appear in tandem. The exercise gives the
right quantum numbers. The Lagrangian for the Glashow--Weinberg--Salam
model, except for the symmetry-breaking terms, is elegantly rewritten
in terms of the Dirac operator only~\cite{ChavesM}. An interesting
fact is that, with the appropriate variables, the Lagrangian for the
gravity field can be written in similar form. Now, the ``dualized
standard models'' try to explain the existence of three generations
and the relations among them from a duality (effectively taking place
at the 100--$1000 \TeV$ scales) between the $\sul(3)$ color sector of
the SM and the fermion families space. This scheme is open to
criticism in that it is unclear why the leptons, which do not see
color, should see dualized color. But the NCG approach taught us that
there is a third~3 in the Standard Model. So, just maybe, there is a
\textit{triality} lurking behind it.

Apart from the above (which is mostly old hat), we would say that the
first stage of the process of linking NCG and fundamental physics
finished around the winter of 1995--96. The humble approach paid
handsomely up, when the very structure of the SM itself gave
Connes the inspiration for the construction of noncommutative spin
manifolds. That is, since that time we know how to put fermion fields
on a noncommutative manifold, and this knowledge is backed by a
fundamental mathematical result, whose proof you can find in
{\it Elements of Noncommutative Geometry\/}~\cite{Polaris}. This needs
in turn the proof of the ``character theorem'', stated but not proved
in~\cite{Book}, which was provided to us by Connes.

Afterwards, there has been little activity on this front. The vexing
question of the unimodularity condition, which has to be imposed on
the noncommutative gauge field~\eqref{eq:action2} in order to really
reproduce the Standard Model with the correct hypercharges, was
``explained'' by Alvarez, Mart\'{\i}n and myself in terms of anomaly
cancellation~\cite{Chiron}; but not understood on a native NCG basis.
This loose thread has been tied up recently by Serge Lazzarini and
Thomas Sch\"ucker~\cite{LazzariniS}.

\section{Arrival of the Top-down Approach}

We have seen that the first cohabitation, in recent times, of
noncommutative geometry and the physics of fundamental interactions
actually followed a \textit{bottom-up} pattern. The \textit{top-down}
approach goes back to the classic paper by Snyder in January 1947,
``Quantized space-time''~\cite{Snyder}.

In this charming paper, it was first suggested that coordinates are
noncommuting operators; the six commutators of the coordinates are
proportional to a basic unit of length and correspond to the
infinitesimal generators of the Lorentz group; throughout, Lorentz
covariance is maintained. Then as now, motivations for using
noncommuting coordinates to describe spacetime were the hope of
improving the renormalizability of QFT and of grappling with the
attendant nonlocality of physics at the Planck scale (uncertainty
relations for spacetime coming from gravity).

The top-down approach has been resurrected recently by string
theorists. In their most popular model, the commutation relations are
simply of the form
\begin{equation}
[x^j, x^k] = i\, \theta^{jk},
\label{eq:comm-reln}
\end{equation}
breaking Lorentz invariance. As anticipated by
Sheikh-Jabbari~\cite{Jabbari} and from a slightly different viewpoint
by Schomerus~\cite{Schomerus}, and plausibly argued by Seiberg and
Witten~\cite{SeibergW}, open strings with allowed endpoints on
2Dirich\-let-branes in a $B$-field background act as electric dipoles
of the abelian gauge field of the brane; the endpoints live on the
noncommutative space determined by~\eqref{eq:comm-reln}. Then
Kar~\cite{Kar} argued that the holographic correspondence relates the
noncommutative and commutative worlds.

Slightly before, Douglas, Schwarz and Connes himself~\cite{ConnesDS}
had arrived at the conclusion that compactification of $M$-theory, in
the context of dimensionally reduced gauge theory actions, leads
ineluctably to noncommutative spaces. And Douglas and Hull~\cite{DouglasH}
had inferred from this that gauge theories on NC tori make sense as a 
limit in string theory. Other work relating noncommutative geometry and
strings includes~\cite{LandiLS}, partly inspired by~\cite{FroehlichG}.

It should also be noted that
Doplicher, Fredenhagen and Roberts had earlier considered a model
with commutation relations
\begin{equation}
[x^j, x^k] = i\, \sigma^{jk},
\label{eq:DFRcomm-reln}
\end{equation}
where the $\sigma^{jk}$ are the components of a tensor, and beyond
that are $c$-numbers for all practical purposes. Thus in their
formalism~\cite{DoplicherFR}, Lorentz invariance is also explicitly
kept.

A very important construction in the framework of~\cite{SeibergW} is
the so-called Seiberg--Witten map in gauge theory. This map relates
the gauge fields and the gauge variations in a noncommutative theory
with commutative counterparts. Let $\Ahat(A)$ and $\lahat(A,\la)$,
where
$$
\hat\delta \Ahat_j
= \del_j \lahat + i (\Ahat_j \star_\theta \lahat - \lahat
\star_\theta \Ahat_j),
$$
be, respectively, the NC gauge potentials and gauge variation in terms
of the commutative ones; the $\star_\theta$ denotes the Moyal product,
of which more below. Then necessarily, in order to have
$$
\Ahat(A) + \hat\delta\,\Ahat(A) = \Ahat(A + \delta A)
$$
for $\theta$ small,
\begin{align}
\Ahat_j(A) 
&= A_j - \tquarter \theta^{kl} \{A_k, \del_l A_j + F_{lj} \}
   + O(\theta^2),
\nonumber \\
\lahat(A,\la)
&= \la + \tquarter \theta^{kl} \{\del_k \la, A_l \} + O(\theta^2),
\label{eq:sw-map}
\end{align}
where $F(A)$ is the ordinary gauge field for~$A$. These equations have
been seen by Jur\v{c}o, Schupp and Wess~\cite{JurcoSW}, and also by
Jackiw and Pi~\cite{JackiwP}, to correspond to an infinitesimal
1-cocycle for a projective representation of the underlying gauge
group in the Moyal algebra.

The equations \eqref{eq:sw-map} are valid for arbitrary $\theta$, and 
so, when using the Seiberg--Witten map, the components of $\theta$ 
should be regarded as variables. 

\medskip

Before continuing, we need an interlude on the Moyal product. It is
correctly (nonperturbatively) defined, for nondegenerate $\theta$, as
\begin{equation}
f \star_\theta g(u) := (\pi\theta)^{-4}
\int_{\R^4}\int_{\R^4}\,f(u + s)g(u + t)\,e^{2is\theta^{-1}t}\,ds\,dt,
\label{eq:Moyal-prodint}
\end{equation}
and this gives rise to the commutation relations~\eqref{eq:comm-reln}.

Mathematically, those are the commutation relations of Quantum
Mechanics, when the reduced Planck constant replaces~$\theta$! From
now on, we neglect to indicate $\theta$ in the star product notation.
The precise relation of~\eqref{eq:Moyal-prodint} to the asymptotic
formula usually given as definition of the Moyal product was spelled
out some time ago in~\cite{Nereid}. Indeed the formula
\eqref{eq:Moyal-prodint} is the basis of the ``fourth'' formalism of
Quantum Mechanics, in which observables, states and transitions are
described by functions (or distributions) on phase space. This
formalism, which goes back to Wigner, Weyl and Moyal, had already a
long and proud history when (a version of) the Moyal product was
rediscovered by string theorists. And so, as Fedele Lizzi has put it,
``in the string community we can happily go on rediscovering quantum
mechanics''~\cite{LizziViewpoint}.

An even more fortunate fact was that the quantum field-theoretical
framework for making sense of the noncommutative limit of string
theory also preexisted the paper by Seiberg and Witten. This story
deserves its own chapter.

\section{Noncommutative Field Theory (NCFT)}

NCFT can be and was developed independently of its string theory
motivation and background. I go not into the theory of quantum fields
on ad-hoc discrete spaces (in this Conference, Rainer H\"au{\ss}ling
will touch on that). In turn the latter topic should not be confused
with ``fuzzy physics'', which has led nowhere, and does deserve a
sharp look ---but not here. A reasonably good review for NCFT
is~\cite{DouglasN}.

Let us explain why NC field theory preexisted the Seiberg--Witten
paper. Connes' noncommutative spin manifold theorem and Fredholm
module theory, referred to above, come in handy here. Quantum field
theory has an algebraic core independent of the nature of space-time.
From the representation theory of the infinite dimensional orthogonal
group, or an appropriate sugbroup thereof, with the input of a
single-particle space, it is possible to derive all Fock space
quantities of interest. Nothing really changes if the ``matter field''
evolves on a noncommutative space. In a nutshell: one can apply the
canonical quantization machinery to a noncommutative kind of
single-particle space~\cite{Atlas}.

We already mentioned the long-standing hope that giving up
\textit{locality} in the interaction of fields ---one of the basic
tenets of quantum field theory and indeed one of the main selling
points by the forefathers--- would be rewarded with a better
ultraviolet behaviour. This hope was now amenable to rigorous
scrutiny, and it is {\it not\/} borne out. QFT on noncommutative
manifolds also \textit{requires renormalization}. This, in some sense
the first result of NCFT, was proved in general by V\'arilly and
myself in~\cite{Atlas}. A subsequent important paper by Chaichian,
Demichev and Pre\v{s}najder, written in a different spirit, confirmed
it~\cite{ChaichianDP}.

Of course, one can prove the same in the context of a
\textit{particular} NCG model, by writing down the integral
corresponding to a Feynman diagram, and finding it to be divergent.
That had been shown previously by Filk~\cite{Filk}, for the scalar
Lagrangian theory associated the Moyal product algebra. (Filk had been
active in ``Twisted Eguchi--Kawai'' models, and retrospectively it
seems clear that he took his cue from~\cite{GonzalezAKA}, where these
models were generalized to continuous momenta.)

Two other pioneering papers which predated the one by Seiberg and
Witten were \cite{MartinSRone} and \cite{KrajewskiW}. These are important
because they show that, at the one-loop level, the radiative
corrections respect the gauge (BRS) invariance of the theory. This is
not trivial in noncommutative Yang--Mills theory: for instance, it
involves curious cancellations for the 4-point function.

Before enumerating the main issues in NCFT, let us issue a warning.
There has been an enormous outpouring of literature concerning those
issues. The quality of many of those papers is substandard; they
should come with an attached notice of \textit{Warnung}. To begin
with, the mathematics of the Moyal product, as already said, antedates
its current use by string theorists. There has been no shortage of
mathematical mistakes, attesting that many people did not bother to
study the Weyl--Wigner--Moyal formalism in depth. For instance,
in~\cite{Harvey} we find the following assertion (unchallenged till
now, to our knowledge):
\begin{quotation}
Let $\SS^m$ be the set of functions on phase space defined by
the condition $\SS^m = \set{f(q,p) : |\del_q^\a \del_p^\b f|
\leq C_{\a\b} (1 + q^2 + p^2)^{(m-\a-\b)/2}}$, for constants
$C_{\a\b}$. Then one can show that $f$ being in $\SS^m$ with
$m \leq 0$ implies that $\hat O_f \in \mathcal{B}(\H)$ while $m < 0$
implies that $\hat O_f \in \K(\H)$. Thus we see that the Weyl
transforms of bounded operators are bounded functions [\dots]
\end{quotation}
The last phrase is a {\it non sequitur\/}. It is moreover incorrect.
For instance, the (inverse) Weyl transform of the parity operator,
which is obviously bounded, is a delta function~\cite{Iapetus}. In
general, the Fourier transform of a function giving a bounded operator
gives a bounded operator, too, although perhaps for a different value
of $\theta$. The correct approach has been in the literature at least
since~\cite{Phobos,Deimos}. Also, some papers read as if they were
conceived by randomly opening a book on QFT, and trying to make the
noncommutative analog of the section which appeared. Particularly
obnoxious is the pest of trivial observations, obvious to anyone
steeped in noncommutative mathematics, being republished again and
again. For example, the note~\cite{ChaichianPJTNogo} just rehashes a
remark in the (substantial) paper~\cite{Terashima}. The whole field
shows the traces of having grown too quickly, tending to obscure the
lasting contributions.

\smallskip
What then are the main issues in NCFT?

\begin{itemize}

\item  The distinction between planar and nonplanar Feynman diagrams.

Envisage, for instance, the theory given by the action functional
$$
S = \int d^4x\, \biggl(\frac{1}{2} \pd{\phi}{x^\mu} \pd{\phi}{x_\mu} 
+ \frac{1}{2} m^2\phi^2 + \frac{g}{4!} \phi\star\phi\star\phi\star\phi 
\biggr)
$$
Generating the Feynman rules like in the commutative theory, the
propagators do not change, but the vertices get in momentum space a
factor proportional to
$$
\exp\biggl(-i/2\sum_{1\leq k<l\leq 4}p_{k\a}\theta_{\a\b}p_{l\b}\biggr);
$$
we suppose all the momenta are incoming on the vertex, in cyclic order.

Consider then a simple diagram of the model like the tadpole diagram;
let $p$ denote the incoming momentum and $k$ the loop variable.
Depending on the order of the momenta, the previous factor is equal
to~1 or to $e^{-ip_\a\theta_{\a\b}k_\b}$.

In general, planar diagrams are those that get overall phase factors
depending only on the external legs; for nonplanar diagrams there are
phase factors which depend on loop variables as well, and the
corresponding integrals become convergent. For the tadpole diagram, we
get amplitudes of the form
$$
\Ga_{\rm pl}(p)\propto \int \frac{d^4k}{k^2 + m^2},  \qquad
\Ga_{\rm npl}(p)\propto \int \frac{d^4k}{k^2 + m^2}\, e^{-ip\theta k}.
$$
The second integral is finite. Actually, the combinatorics of the
nonlocal vertex factors was worked out long ago by Gonz\'alez-Arroyo
and Okawa in the context of the TEK models, and given in detail
in~\cite{KorthalsAltes}.

For noncommutative Yang--Mills theory at the one-loop level, have a
look at~\cite{MartinSRtwo}.

\item  The UV/IR mixing and supersymmetry.

Nonplanar diagrams may become divergent again for particular values of
the momenta. This would not seem to matter as long as
$\Ga_{\rm npl}(p)$ remains a well defined distribution. But, for more
complicated diagrams with subdivergences, the dependence on~$p$ of
their amplitude behaviour spells trouble, because these diagrams may
unexpectedly become divergent again. This is the notorious UV/IR
mixing, which tends to spoil renormalizability. One can think it
reflects the underlying ``stringy degrees of freedom''. For Moyal
noncommutative Yang--Mills theory, this happens already at the 2-loop
level. People have tried to get around the obstacle using resummation
techniques, or nonlocal field redefinitions, but none of those are
convincing to this reviewer so far.

Supersymmetric theories have advantages in regard to renormalizability
and the UV/IR trouble~\cite{GirottiGRS}. In this context, we remark it
has been proved by Paban, Sethi and Stern~\cite{PabanSS} that the
deformation leading to NC Yang--Mills theory in the sense of Seiberg
and Witten is about the only one compatible with supersymmetry.

\item  Renormalizability in the context of the Seiberg--Witten map.

Another possibility in order to circumvent the UV/IR problem and
obtain renormalizable theories is to use the Seiberg--Witten map. The
trouble with this approach is that many new vertices appear, at
different orders in $\theta$ and~$\hbar$. Extensive
calculations~\cite{WulkenhaarTheta} have shown that in
$\theta$-expanded QED, many miraculous cancellations occur. However,
the fermion 4-point function becomes divergent, so, strictly speaking,
that model is nonrenormalizable.

Since in this Conference Raimar Wulkenhaar talks about the subject, I
shall not spend more time on these matters.

\item  The question of non/unitarity of theories on spaces with
timelike noncommutativity.

It has been asserted~\cite{GomisM} that theories with timelike
noncommutativity suffer from violation of unitarity. The issue was
lucidly discussed in~\cite{LAGBarbonZ}. However, Dorothea Bahns
et~al~\cite{BahnsDFP} have cogently argued that, when using a
Hamiltonian approach to NCFT, instead of the usual Lagrangian
approach, this problem can be exorcised. The change in viewpoint
concerns only the nonplanar diagrams. 

(As we were preparing the manuscript for publication, the idea by
Bahns et~al has been further elaborated by~\cite{RimY}
and~\cite{LiaoS}. We look forward to hear more on this.)

In this context, investigation of the old model by Snyder, unjustly
neglected in our opinion, is probably worthwhile. The Hamiltonian
approach, nevertheless, probably does not help with UV/IR
trouble~\cite{BahnsPrivate}.

\item  The construction of gauge-covariant/invariant observables.

It will have been clear to the listener that the space coordinates 
themselves are not gauge-covariant. On the other hand, 
$x^\mu + \theta^{\mu\nu} A_\nu$ is a gauge-covariant quantity
---as was well known to the
practitioners of Weyl--Wigner--Moyal theory. This noncovariance of the
space coordinates brings NC theories closer with gravity. The
construction of gauge-invariant observables has been competently taken
up by Harald Dorn and coworkers in a series of papers~\cite{Dorn}.

\item  Anomalies in NCFT.

The part of the nonabelian anomaly which is quadratic in the gauge
potentials was found by Carmelo P. Mart\'{\i}n and
myself~\cite{Camilla} to be
\begin{align*}
&-\frac{1}{96\pi^2}\,\int d^4 x\,
\eps_{\mu_1\mu_2\mu_3\mu_4}\,\Tr\,T^a\,[T^b,T^c]\,
\del_{\mu_1} \theta^a \bigl[
\Ahat^b_{\mu_2} \star \del_{\mu_3} \Ahat^c_{\mu_4} -
\del_{\mu_3} \Ahat^c_{\mu_4} \star \Ahat^b_{\mu_2} \bigr]
\\
&-\frac{1}{96\pi^2}\,\int d^4 x\,
\eps_{\mu_1\mu_2\mu_3\mu_4}\,\Tr\,T^a\,\{T^b,T^c\}\,
\del_{\mu_1} \theta^a \bigl[
\Ahat^b_{\mu_2} \star \del_{\mu_3} \Ahat^c_{\mu_4} +
\del_{\mu_3} \Ahat^c_{\mu_4} \star \Ahat^b_{\mu_2} \bigr].
\end{align*}
The first term of this innocent-looking formula is a new contribution
that does not vanish in the noncommutative case. It makes it quite
difficult, albeit not impossible~\cite{IntriligatorK}, to construct
nonanomalous chiral theories in NCG.

Further work by C.~P. Mart\'{\i}n~\cite{MartinOrigin} has established
that the anomalies are related to the (divergent) planar diagrams:
nonplanar diagrams would not break the Ward identities. This last
contention, however, has been called in question by~\cite{ArmoniLT}.

\end{itemize}

In summary, NCFT offers a brand-new laboratory for QFT. However, when 
eating this fish, one should beware of the fishbones.

\section{Phenomenological Window}

A game one can now play is to marry the Connes--Lott approach to the
new commutation relations. That is to say, to work with the tensor
product of (some variant of) the Eigenschaften algebra and a spacetime
algebra which is no longer commutative. This has been done, for
instance, in the following papers: \cite{Morita}, by K. Morita;
\cite{ChaichianPJTModel}, by Chaichian et~al; and, perturbatively in
$\theta$, in~\cite{CalmetJSWW}, by the M\"unchen group. These models
should be checked for anomalies.

Noncommutative spacetime coordinates lead to signatures in several
$2 \to 2$ QED tree--level processes in $e^+e^-$ collisions which are
in principle observable. In particular, corrections to the amplitudes
for pair annihilation, M{\o}ller and Bhabha scattering, as well as
$\ga\ga \to \ga\ga$ scattering, for noncommutative scales
$\La_{\rm NC}$ of order $\geq 1 \TeV$, have been calculated this way
and could in principle be probed at linear colliders~\cite{Hewett}.
(Here, $\La_{\rm NC}$ is defined as $1/\sqrt{\theta}$, where $\theta$
is a typical parameter in the Moyal product algebra.) Naturally, the
wisdom of these calculations, in absence of a proper interpretation of
the model at the 2-loop level, is somewhat questionable.

There have also been discussions on the signature of noncommutativity
in CP violation phenomena, and the muon magnetic moment ``anomaly''.

The cosmological implications of a nonzero noncommutativity parameter
are also open to discussion. Some people~\cite{ChuGS,LizziMMP} feel
that it will modify the quantum perturbations of the inflaton field;
it does not seem that noncommutativity can do away with inflation
altogether.

\section{Spectral Action and Related Developments}

Meanwhile, Alain Connes has not remained idle. Dissatisfaction with
the absence of gravity in the old Connes--Lott model led him to
develop the spectral action principle. Together with
Chamseddine~\cite{ChamseddineCSpec}, he proposed a universal formula
for an action associated with a (compact) noncommutative spin geometry. The
action is based on the spectrum of the Dirac operator and is a
geometric invariant. Automorphisms of the manifold underlying the
geometry combine both diffeomorphisms and internal symmetries.

The Yang--Mills action functional is there replaced by a ``universal''
bosonic functional of the form:
$$
B_\phi[D] = \Tr \phi(D^2),
$$
with $\phi$ being an ``arbitrary'' positive function of the Dirac
operator~$D$. Chamseddine and Connes argue that $B_\phi$ has the
following asymptotic development:
$$
B_\phi[D/\La] \sim \sum_{n=0}^\infty f_n\, \La^{4-2n}\, a_n(D^2)
\as \La \to \infty,
$$
where the $a_n$ are the coefficients of the heat kernel expansion for
$D^2$ and $f_0 = \int_0^\infty x \phi(x) \,dx$,
$f_1 = \int_0^\infty \phi(x) \,dx$, $f_2 = \phi(0)$,
$f_3 = -\phi'(0)$, and so on. This was put on a sounder mathematical
footing in~\cite{Odysseus}.

They proceed to compute the development for the Dirac--Yukawa operator
associated to the Standard Model, obtaining all terms in the bosonic
part of the action for the Standard Model, plus unavoidable gravity
couplings. That is to say, the spectral action for the Standard Model
unifies with gravity at a very high energy scale. Later, a variant of
the spectral action was suggested in~\cite{Zappafrank}.

Wulkenhaar has conjectured that the spectral action for the
Moyal algebra has the necessary additional symmetries for gauge
theories on $\theta$-deformed spacetime to become renormalizable. To
check this, one needs to extend the spectral action to the noncompact
NC manifold case; this is indeed one of the burning questions of
the hour~\cite{Selene}.

\medskip

Somewhat related to the above, Connes~\cite{ConnesSurvey} and also
Landi~\cite{ConnesLa}, V\'arilly~\cite{Larissa},
Sitarz~\cite{SitarzSphere} and Dubois-Violette~\cite{ConnesDV}, have
been working on systematically obtaining new noncommutative spaces.
One motivation for that is to do quantum gravity in the noncommutative
Euclidean context. For that, commutative and noncommutative manifolds
alike should be ``counted'' somehow. The difficulties to count
4-manifolds in the context for quantum gravity are well known ---see
the magisterial account by E. Alvarez in~\cite{AlvarezQgrav}.
Algebraic $K$-theory maybe offers a tool to bypass the problem. Note
that one should consider ``virtual'' NC manifolds; fulfilment of the
Hochschild condition~\cite{ConnesGrav} might be the criterion to
distinguish true classical NC manifolds from virtual
ones~\cite{ConnesRemarkOW}.

\section{On the Connes--Kreimer Hopf Algebra}

Bogoliubov's combinatorial renormalization scheme in dimensional
regularization can be summarized as follows. If $\Ga$ is 1PI and is
\textit{primitive} (i.e., it has no subdivergences), set
$$
C(\Ga) := - T(f(\Ga)),  \sepword{and then}  R(\Ga) := f(\Ga) + C(\Ga),
$$
where $C(\Ga)$ is the \textit{counterterm} and $R(\Ga)$ is the desired
finite value: in other words, for primitive graphs one simply removes
the pole part. Next, we may recursively define Bogoliubov's
$\Rbar$-operation by setting
$$
\Rbar(\Ga) = f(\Ga)
 + \sum_{\emptyset\subsetneq\ga\subsetneq\Ga} C(\ga)\,f(\Ga/\ga),
$$
with the proviso that
\begin{equation}
C(\ga_1\dots\ga_r) := C(\ga_1) \dots C(\ga_r),
\label{eq:mult-ctrtm}
\end{equation}
whenever $\ga = \ga_1\dots\ga_r$ is a disjoint union of several
components. The final result is obtained by removing the pole part of
the previous expression: $C(\Ga) := - T(\Rbar(\Ga))$ and
$R(\Ga) := \Rbar(\Ga) + C(\Ga)$. In summary,
\begin{subequations}
\label{eq:graph-renorm}
\begin{align}
C(\Ga) &:= - T\biggl[ f(\Ga)
+ \sum_{\emptyset\subsetneq\ga\subsetneq\Ga}C(\ga)\,f(\Ga/\ga)\biggr],
\label{eq:graph-renorm-C}
\\
R(\Ga) &:= f(\Ga) + C(\Ga)
 + \sum_{\emptyset\subsetneq\ga\subsetneq\Ga} C(\ga)\,f(\Ga/\ga).
\label{eq:graph-renorm-R}
\end{align}
\end{subequations}

Let now $\Phi$ stand for any particular QFT. The Hopf algebra $H_\Phi$
is a commutative algebra generated by one-particle irreducible graphs:
that is, connected graphs with at least two vertices which cannot be
disconnected by removing a single line. The product $\Ga_1 \Ga_2$
means the disjoint union of the graphs $\Ga_1$ and $\Ga_2$. The counit
is given by $\eps(\Ga) := 0$ on any generator, with
$\eps(\emptyset) := 1$ (we assign the empty graph to the identity
element). The \textit{coproduct} $\Dl$ is given, on any 1PI graph
$\Ga$, by
$$
\Dl\Ga := \sum_{\emptyset\subseteq\ga\subseteq\Ga} \ga \ox \Ga/\ga,
$$
where the sum ranges over all subgraphs which are divergent and proper
(in the sense that removing one internal line cannot increase the
number of its connected components); $\ga$ may be either connected or
a disjoint union of several connected pieces. The terms for
$\ga = \emptyset$ and $\ga = \Ga$ in the sum are
$\Ga \ox 1 + 1 \ox \Ga$. The notation $\Ga/\ga$ denotes the
(connected, 1PI) graph obtained from~$\Ga$ by replacing each component
of $\ga$ by a single vertex.

To see that $\Dl$ is coassociative, we observe that if
$\ga \subseteq \ga' \subseteq \Ga$, then $\ga'/\ga$ can be regarded as
a subgraph of $\Ga/\ga$; moreover, it is obvious that
$(\Ga/\ga)/(\ga'/\ga) \simeq \Ga/\ga'$. The desired relation
$(\Dl \ox \id)(\Dl\Ga) = (\id \ox \Dl)(\Dl\Ga)$ can now be expressed
as
$$
\sum_{\emptyset\subseteq\ga\subseteq\ga'\subseteq\Ga}
 \ga \ox \ga'/\ga \ox \Ga/\ga'
 = \sum_{\emptyset\subseteq\ga\subseteq\Ga,\;
         \emptyset\subseteq\ga''\subseteq\Ga/\ga}
    \ga \ox \ga'' \ox (\Ga/\ga)/\ga'',
$$
which is easy to verify directly.

Here (Figures~\ref{fig:sunset}--\ref{fig:ragdoll}) are some
\textit{coproducts} for $\varphi_4^4$ diagrams:

\begin{figure}[htb]
\centering
$$
\Dl \Bigl( \sunset \Bigr) = \1 \ox \sunset + \sunset \ox \1
$$
\caption{The ``setting sun'': a primitive diagram}
\label{fig:sunset}
\end{figure}

\begin{figure}[htb]
\centering
$$
\Dl \Bigl( \iice \Bigr) = 
\1 \ox \iice + 2\ \sweet \ox \sunset \put(-35,0){$\bullet$}
+ \iice \ox \1
$$
\caption{The ``double ice cream in a cup''}
\label{fig:icecream}
\end{figure}

\begin{figure}[htb]
\centering
$$
\Dl \Bigl(\,\,\, \tsweet \Bigr) = \1 \ox \,\tsweet +
2\ \sweet \ox \,\dsweet\put(-46.5,0){$\bullet$} +
\,\sweet \ox\ \dsweet\put(-26,0){$\bullet$}
$$
$$
+ 2\,\ \dsweet \ox \,\sweet\put(-32,0){$\bullet$} + \,
\sweet\ \sweet \ox \,\sweet\put(-32,0){$\bullet$}
\put(-12,0){$\bullet$} + \ \tsweet\, \ox \1
$$
\caption{The ``triple sweet''}
\label{fig:trisweet}
\end{figure}

\begin{figure}[htb]
\centering
$$
\Dl \Bigl(\,\,\, \kg \Bigr) = \1 \ox \,\kg + \
\sweet \ox \pinterIV\put(-39,0){$\bullet$} + \
\sweet \ox \,\dsweet\put(-6,0){$\bullet$}
$$
$$
+ \ \pinterIV \ox \,\sweet\put(-12,0){$\bullet$} + \
\sweet\ \sweet \ox \,\sweet\put(-32,0){$\bullet$}
\put(-12,0){$\bullet$} + \ \kg \ox \1
$$
\caption{The ``rag-doll''}
\label{fig:ragdoll}
\end{figure}

That defines $H_\Phi$ as a bialgebra. A {\it grading\/} is provided by
\textit{depth}. A graph $\Ga$ in the space $H_\Phi$ has depth $k$ (or
is $k$-primitive) if
$$
P^{\otimes k+1}(\Dl^k\Ga) = 0  \sepword{and}
P^{\otimes k}(\Dl^{k-1})\Ga \neq 0.
$$
where $P$ is the projection $u\circ\eps - \id$, whose importance was
discovered ``experimentally'' by Figueroa and myself~\cite{Ananke} in
the context of the Hopf algebra of rooted trees~\cite{ConnesKrHopf}.

The depth of any graph is finite. In our context, depth measures the
maximal length of the inclusion \textit{chains} of subgraphs appearing
in the Bogoliubov recursion. If $\ga\subsetneq\Ga$ and $\Ga$ has
depth~$l$, then $\ga$ has depth $l-1$ at most. Elements of $H_\Phi$
without subdivergences have depth~1, obviously. This is the case of
the ``fish'' graph in the $\varphi^4_4$ model. The ``ice-cream'' graph
has depth~2. Of the eight four-vertex (three-loop) graphs relevant for
the four-point structure of that model, five have depth~3, two have
depth~2 and one (the tetrahedron graph) has depth~1. In dimensional
regularization, a graph of depth $l$ is expected to display a pole of
$l$th order.

Now $S$ can also be defined as the inverse of $\id = u\circ\eps - P$
for the convolution, and so (after some calculation)
\begin{equation}
S(\Ga_l) := \sum_{k=1}^l P^{*k}\,\Ga_l =
-\Ga_l + \sum_{\emptyset\subsetneq\ga\subsetneq\Ga_l}S(\ga)\,\Ga_l/\ga,
\label{eq:graph-antp}
\end{equation}
for any graph $\Ga_l$ of depth $l$.

The foregoing is my own summary of the Hopf algebra of Feynman
diagrams discovered by Connes and
Kreimer~\cite{ConnesKrRHI,ConnesKrRHII}. As it stands, the Hopf
algebra $H_\Phi$ corresponds to a formal manipulation of graphs. These
formulas can then be matched to expressions for numerical values.
Firstly, the Feynman rules for the unrenormalized theory prescribe a
linear map
$$
f : H_\Phi \to \A
$$
into some commutative algebra $\A$, which is multiplicative on
disjoint unions: $f(\Ga_1 \Ga_2) = f(\Ga_1)\,f(\Ga_2)$. In other
words, $f$ is actually a homomorphism of algebras. In dimensional
regularization, $\A$ is an algebra of Laurent series in a complex
parameter~$\eps$, and $\A$ is the direct sum of two
\textit{subalgebras}:
$$
\A = \A_+ \oplus \A_-.
$$
Let $T\: \A \to \A_-$ be the projection on the second subalgebra, with
$\ker T = \A_+$. $\A_+$ is the holomorphic subalgebra of Taylor series
and $\A_-$ is the subalgebra of polynomials in $1/\eps$ without
constant term; the projection $T$ picks out the pole part, in a
minimal subtraction scheme. Now $T$ is not a homomorphism, but the
property that both its kernel and image are subalgebras is reflected
in a ``multiplicativity constraint'':
$T(ab) + T(a)\,T(b) = T(T(a)\,b) + T(a\,T(b))$.

The equation \eqref{eq:graph-renorm-C} means that ``the antipode
delivers the counterterm'': one replaces $S$ in the calculation
\eqref{eq:graph-antp} by~$C$ to obtain the right hand side, before
projection with~$T$. From the definition of the coproduct in $H_\Phi$,
\eqref{eq:graph-renorm-R} is a \textit{convolution} in
$\Hom(H_\Phi,\A)$, namely, $R = C * f$. To show that $R$ is
multiplicative, it is enough to verify that the counterterm map $C$ is
multiplicative, since the convolution of homomorphisms is a
homomorphism because $\A$ is commutative. In other words, one must
check that \eqref{eq:mult-ctrtm} and \eqref{eq:graph-renorm-C} are
compatible; this was done by Connes and Kreimer in~\cite{ConnesKrRHI},
and elaborated or exploited in subsequent papers; see~\cite{GirelliMK}
in this connection.

The previous discussion is \textit{logically independent} of NCG,
although it is true that the Hopf algebra approach to renormalization
theory has been developed in close contact with the Connes--Moscovici
noncommutative theory of foliations. In that theory, a foliation is
described by a noncommutative algebra of functions twisted by local
diffeomorphisms, and horizontal and vertical vector fields are
represented on that algebra by the action of a certain Hopf algebra
$H_{CM}$ which provides a way to compute a local index
formula~\cite{ConnesMHopf}.

On extending the Hopf algebra $H_\Phi$ of
graphs by incorporating operations of insertion of subgraphs, one
obtains a noncommutative Hopf algebra of the $H_{CM}$ type, which
gives a supplementary handle on the combinatorial structure of
$H_\Phi$~\cite{ConnesKrLie}.

\smallskip
When handled through the Hopf algebra of Feynman
diagrams $H_\Phi$ of Connes and Kreimer, the Epstein--Glaser (EG)
renormalization method provides an alternative route to the core
results of renormalization theory~\cite{PinterSmat,PinterHopf}.

In the last few years, EG renormalization has been revived by the
Hamburg group around Fredenhagen. It has also been sharpened into a
more versatile tool (``$T$-renormali\-za\-tion'') in my~\cite{Carme}.
The concept of \textit{optimal scaling} of the amplitudes is the soul
of this improvement. A similar tack was recently taken by Hollands and
Wald~\cite{HollandsW} for the precise definition of local, covariant
quantum field. See also, in this
respect~\cite{BrunettiFV,JunkerS,Grigore}. The basic idea (very much
in the spirit of NCG) is that, if a quantum field is thought of as a
distribution on a globally hyperbolic spacetime, with values in the
algebra of the Wick products, then isometric embeddings from one
spacetime into another are realized as suitable homomorphisms of the
fields. Then, the key property in the existence proof of time-ordered
products by Hollands and Wald is the postulated, and then recursively
proved, optimal rescaling of functionals of the metric $g$, up to the
logarithmic terms:
$$
\la^d f_n[\la^2 g] = f_n[g] + \log\la \,f_{n-1}[g].
$$

The extension to curved backgrounds of the Connes--Kreimer Hopf
algebra approach to renormalization seems to hinge on the EG method.

\section{Outlook}

Will NCG live up to its promise with regard to fundamental physics? 

One should not confuse the learning of a new language with the
expansion of physical knowledge. A word of caution about the changing
relationships between physics and mathematics is always in order. For
instance, Yang and Mills found a family of theories which seemed
pertinent to describe the observed phenomena. Mathematicians have
translated them into fibre bundle language, and now we teach
Yang--Mills theories to our students with the rigidity of formalism,
so they seem completely unavoidable. However, in their present form
they are suspect (at least to Mills~\cite{Mills} and to this
reviewer), for the good reason that quantizing them is a nightmare. It
may come to pass that eventually they be seen as an incomplete
preliminary stage of a more refined physical model.

By the standards of fashion, the NCG conquest of physics is now ebbing
somewhat. This ebb will surely prove temporary, since the concepts and
tools of noncommutative geometry are indeed powerful and helpful in
QFT. On the other hand, our ability to reformulate the SM as a
noncommutative geometry is not, in itself, all that meaningful. A
truer test for the NCG paradigm is the understanding of the fermion
mass and mixing matrices.

As for NC field theory, the models available so far, while they amount
to a useful mathematical laboratory for the unification of gravity and the
other fundamental interactions, are still too rough. In spite of the
ideology of effective field theory, it remains true that the fantastic
successes of the renormalizability program, now validated for
geometric bakgrounds, and of renormalizability itself as a heuristic
principle in selecting physical models, tell us that the violation of
locality which we expect at some level is subtler than we have been
able to dream of until now. Some crucial ingredient is missing.

\subsection*{Acknowledgements}

I am very grateful to the organizers of the 66.~Tagung der Deutsche
Physikalische Gesellschaft for the invitation to survey the field.
Financial support for my stay at the Departamento de F\'{\i}sica
Te\'orica I of Universidad Complutense came from the Secretar\'{\i}a
de Estado de Educaci\'on y Universidades of Spain. Carmelo P.
Mart\'{\i}n and Joseph C. V\'arilly have been helpful, and influential
in shaping the opinions outlined here. Serge Lazzarini taught me the
EG scheme. Special thanks are due to Roger Plymen and Henri
Moscovici for sharing with me recollections of the early days of
noncommutative geometry.

\end{document}